\documentclass[prl, twocolumn]{revtex4-1}
	
	\usepackage{graphicx}
	\usepackage{color}
	\usepackage{verbatim}

	\usepackage{subfigure}
	\usepackage{amssymb}
	\usepackage{amsmath}
	\usepackage[normalem]{ulem}  % added this so I can use \sout to strikeout words in the text
	
	\newcommand{\dm}[1]{{\color{black}#1}}

\begin{document}
	
	\title{Reply to ``Comment on \\ `Entropy production and fluctuation theorems for active matter' ''}
	\author{Dibyendu Mandal$^{1}$ and Katherine Klymko$^{2}$ and Michael R. DeWeese$^{1, 3}$}
	\affiliation{
	$^1$Department of Physics, University of California, Berkeley, CA 94720, USA \\
	$^2$Department of Chemistry, University of California, Berkeley, CA 94720, USA \\
	$^3$Redwood Center for Theoretical Neuroscience and Helen Wills Neuroscience Institute, University of California, Berkeley, CA 94720, USA}
	
	\maketitle

In the preceding Comment~\cite{caprini2018}, Caprini \textit{et al.} criticize our formula for entropy production for active Ornstein-Uhlenbeck processes (AOUP)~\cite{mandal2017entropy}. 
In this Reply, we justify our formula and point out some shortcomings of their proposal. 
	
In our Letter~\cite{mandal2017entropy}, we introduced auxiliary momentum variables to map the original overdamped AOUP to an underdamped Langevin system.
This was also done in Refs.~\cite{fodor2016far, Marconi2017, Puglisi2017}. 
Caprini and colleagues state that AOUP can be analyzed without introducing these auxiliary variables. 
They employ a path-integral description of the colored noise in AOUP and then derive an expression for entropy production that agrees with Refs.~\cite{fodor2016far, Marconi2017} except for a boundary term. 
While we have no concerns about their path-integral treatment, we disagree with their definition of entropy production, $\Sigma \left[\Gamma \right] = \log {(P[\Gamma] / P[\Gamma^r])}$, in the current context. 
This definition leads to the unphysical result of zero entropy production rate in situations where a positive rate is warranted, as previously noted~\cite{fodor2016far, mandal2017entropy}. 
Our work does not suffer from this deficiency~\cite{mandal2017entropy}.  
\dm{While the above} formula is \dm{valid} for overdamped systems with Gaussian white noise, \dm{its usage for AOUP requires an explicit justification that has not been provided in the Comment.}  
Contrary to \dm{this arbitrary assumption}, our definition \dm{ of entropy production} [Eq.~(8) in Ref. \cite{mandal2017entropy}]  is based on a microscopic derivation of the associated Langevin equation \dm{and its time reversal} (see Appendix III in \cite{mandal2017entropy}). 
This allowed us to derive consistent formulations of both the first and second laws of thermodynamics.

\dm{To elaborate}, the above definition of entropy production can be justified within the stochastic energetic framework~\cite{sekimoto_2010} in only two situations: (i) overdamped Langevin systems in contact with an equilibrium reservoir~\cite{Seifert_2012} and (ii) underdamped Langevin systems in contact with an equilibrium reservoir provided there is no momentum-dependent force on the system other than damping.
AOUP does not satisfy these criteria because its bath is out of equilibrium. 
Following Ref.~\cite{fodor2016far}, we therefore introduced auxiliary momentum variables to map AOUP to an underdamped Langevin system that can be interpreted as being in contact with an equilibrium reservoir. 
In the mapped system, there is a momentum-dependent force in addition to the damping force.
As has been shown over the years~\cite{andrieux2008quantum, pradhan2010diamagnetism, chun2018microscopic}, fluctuation relations for entropy production in this situation involve a new process that arises from the reversal of this additional momentum-dependent force. 
We have followed this established procedure in our work. 

%We believe that the authors of the Comment must have misunderstood our use of the word ``auxiliary'' \dm{when we introduced $\{\mathbf{p}_i\}$ as auxiliary momentum variables (just after Eq.~(3b) in Ref.~\cite{mandal2017entropy}}).  
% The authors of the Comment must have misunderstood 
One issue raised in the Comment misrepresented
%our use of the word ``auxiliary'' when we introduced $\{\mathbf{p}_i\}$ as auxiliary momentum variables (just after Eq.~(3b) in Ref.~\cite{mandal2017entropy}).
the nature of the auxiliary variables  $\{\mathbf{p}_i\}$ we introduced after Eq.~(3b) in Ref.~\cite{mandal2017entropy}.
To quote our Letter (with emphasis added) \cite{mandal2017entropy}: ``the overdamped AOUP can be \emph{mapped} exactly to an underdamped Langevin process where the new, \emph{effective} medium (reservoir) is in equilibrium.''
% which clearly conveys the idea that we are using this mapping to constructively solve an otherwise intractable problem in stochastic thermodynamics. 
%We did not interpret the 
This auxiliary momentum is not
%to be 
equivalent to the momentum of the original system.

The authors assert that our detailed fluctuation relation [Eq. (12), Ref.~\cite{mandal2017entropy}] cannot be experimentally verified because it requires measuring $P^r(\Sigma^r)$, which involves ``the probability of entropy production according to a different dynamics, Eq. (7), which is not known to represent any realizable system.'' 
%We take this as encouragement to develop new experiments to validate our results. 
The authors do not provide any specific reason for why they think the reverse process [Eq.~(7), Ref.~\cite{mandal2017entropy}] is not physically realizable. 
In any case, we believe that it is uncontroversial that one should correctly solve the relevant equations for a given theory and then test it with the relevant experiments, even if those measurements are more technically challenging than those needed to test a competing theory --- 
the difficulty of the experiment does not dictate which theory is correct.

Finally, Caprini and colleagues 
%raise the point 
state 
that the AOUP model itself is imperfect in that it neglects the thermal bath altogether, and is only a coarse-grained description of the process of interest. 
Of course, one must typically choose a reduced model in order to study a complex physical system.
Despite being a coarse-grained model, AOUP does not satisfy the fluctuation-dissipation relation and, thus, must possess a positive entropy production rate in its stationary state. 
This is guaranteed in our work but not in the works suggested in the Comment~\cite{fodor2016far, Marconi2017}. 
Caprini and colleagues also raise the concern that, according to our approach, the average heat exchange and entropy production are zero if the external potential is removed ($\Phi = 0$), even if the bath is out of equilibrium ($\tau \neq 0$). 
This they interpret to be a shortcoming of our identification of $-p/(\mu m)+\sqrt{2/(\mu \beta)}\eta$ with a thermal bath. 
%We disagree with this interpretation. 
%When 
In fact, if 
the potential is removed, AOUP is mathematically equivalent to an equilibrium underdamped process,
which should indeed have zero average heat exchange and entropy production.
%and thus it is not a surprise that the average heat exchange and entropy production are zero. 
%We would also like to point out to the authors that we have already taken the first steps towards understanding the more realistic case in our Appendix VIII~\cite{mandal2017entropy}, which discusses an extension of our results when thermal forces are explicitly included.  
%It will be interesting in future studies to apply our rigorous analysis to other model systems of active matter.

\emph{Acknowledgements}:  M.R.D. is grateful for support from the National Science Foundation through Grant No. IIS-1219199. This material was based upon work supported in part by the US Army Research Laboratory and the US Army Research Office under Contract No. W911NF-13-1-0390.

%\bibliography{active.bib}

\begin{thebibliography}{9}%
\makeatletter
\providecommand \@ifxundefined [1]{%
 \@ifx{#1\undefined}
}%
\providecommand \@ifnum [1]{%
 \ifnum #1\expandafter \@firstoftwo
 \else \expandafter \@secondoftwo
 \fi
}%
\providecommand \@ifx [1]{%
 \ifx #1\expandafter \@firstoftwo
 \else \expandafter \@secondoftwo
 \fi
}%
\providecommand \natexlab [1]{#1}%
\providecommand \enquote  [1]{``#1''}%
\providecommand \bibnamefont  [1]{#1}%
\providecommand \bibfnamefont [1]{#1}%
\providecommand \citenamefont [1]{#1}%
\providecommand \href@noop [0]{\@secondoftwo}%
\providecommand \href [0]{\begingroup \@sanitize@url \@href}%
\providecommand \@href[1]{\@@startlink{#1}\@@href}%
\providecommand \@@href[1]{\endgroup#1\@@endlink}%
\providecommand \@sanitize@url [0]{\catcode `\\12\catcode `\$12\catcode
  `\&12\catcode `\#12\catcode `\^12\catcode `\_12\catcode `\%12\relax}%
\providecommand \@@startlink[1]{}%
\providecommand \@@endlink[0]{}%
\providecommand \url  [0]{\begingroup\@sanitize@url \@url }%
\providecommand \@url [1]{\endgroup\@href {#1}{\urlprefix }}%
\providecommand \urlprefix  [0]{URL }%
\providecommand \Eprint [0]{\href }%
\providecommand \doibase [0]{http://dx.doi.org/}%
\providecommand \selectlanguage [0]{\@gobble}%
\providecommand \bibinfo  [0]{\@secondoftwo}%
\providecommand \bibfield  [0]{\@secondoftwo}%
\providecommand \translation [1]{[#1]}%
\providecommand \BibitemOpen [0]{}%
\providecommand \bibitemStop [0]{}%
\providecommand \bibitemNoStop [0]{.\EOS\space}%
\providecommand \EOS [0]{\spacefactor3000\relax}%
\providecommand \BibitemShut  [1]{\csname bibitem#1\endcsname}%
\let\auto@bib@innerbib\@empty
%</preamble>
\bibitem [{\citenamefont {Caprini}\ \emph {et~al.}(2018)\citenamefont {Caprini},
  \citenamefont {Marconi},\citenamefont {Puglisi},\ and\ 
  \citenamefont {Vulpiani}}]{caprini2018}%
  \BibitemOpen
  \bibfield  {author} {\bibinfo {author} {\bibfnamefont {L.}~\bibnamefont
  {Caprini}}, \bibinfo {author} {\bibfnamefont {U.~M.~B.}~\bibnamefont {Marconi}},
  \bibinfo {author} {\bibfnamefont {A.}~\bibnamefont {Puglisi}}, \
  and\ \bibinfo {author} {\bibfnamefont {A.}\ \bibnamefont {Vulpiani}},\
  }\href@noop {} {\bibfield  {journal} {\bibinfo  {journal} {preceding Comment, Physical Review
  Letters}\ }\textbf {\bibinfo {volume} {121}},\ \bibinfo {pages} {LPK1115}
  (\bibinfo {year} {2018})}\BibitemShut {NoStop}%
 \bibitem [{\citenamefont {Mandal}\ \emph {et~al.}(2017)\citenamefont {Mandal},
  \citenamefont {Klymko},\ and\ \citenamefont {DeWeese}}]{mandal2017entropy}%
  \BibitemOpen
  \bibfield  {author} {\bibinfo {author} {\bibfnamefont {D.}~\bibnamefont
  {Mandal}}, \bibinfo {author} {\bibfnamefont {K.}~\bibnamefont {Klymko}}, \
  and\ \bibinfo {author} {\bibfnamefont {M.~R.}\ \bibnamefont {DeWeese}},\
  }\href@noop {} {\bibfield  {journal} {\bibinfo  {journal} {Physical Review
  Letters}\ }\textbf {\bibinfo {volume} {119}},\ \bibinfo {pages} {258001}
  (\bibinfo {year} {2017})}\BibitemShut {NoStop}%
\bibitem [{\citenamefont {Fodor}\ \emph {et~al.}(2016)\citenamefont {Fodor},
  \citenamefont {Nardini}, \citenamefont {Cates}, \citenamefont {Tailleur},
  \citenamefont {Visco},\ and\ \citenamefont {van Wijland}}]{fodor2016far}%
  \BibitemOpen
  \bibfield  {author} {\bibinfo {author} {\bibfnamefont {{\'E}.}~\bibnamefont
  {Fodor}}, \bibinfo {author} {\bibfnamefont {C.}~\bibnamefont {Nardini}},
  \bibinfo {author} {\bibfnamefont {M.~E.}\ \bibnamefont {Cates}}, \bibinfo
  {author} {\bibfnamefont {J.}~\bibnamefont {Tailleur}}, \bibinfo {author}
  {\bibfnamefont {P.}~\bibnamefont {Visco}}, \ and\ \bibinfo {author}
  {\bibfnamefont {F.}~\bibnamefont {van Wijland}},\ }\href@noop {} {\bibfield
  {journal} {\bibinfo  {journal} {Physical Review Letters}\ }\textbf {\bibinfo
  {volume} {117}},\ \bibinfo {pages} {038103} (\bibinfo {year}
  {2016})}\BibitemShut {NoStop}%
\bibitem [{\citenamefont {Marconi}\ \emph {et~al.}(2017)\citenamefont
  {Marconi}, \citenamefont {Puglisi},\ and\ \citenamefont
  {Maggi}}]{Marconi2017}%
  \BibitemOpen
  \bibfield  {author} {\bibinfo {author} {\bibfnamefont {U.~M.~B.}\
  \bibnamefont {Marconi}}, \bibinfo {author} {\bibfnamefont {A.}~\bibnamefont
  {Puglisi}}, \ and\ \bibinfo {author} {\bibfnamefont {C.}~\bibnamefont
  {Maggi}},\ }\href@noop {} {\bibfield  {journal} {\bibinfo  {journal}
  {Scientific Reports}\ }\textbf {\bibinfo {volume} {7}},\ \bibinfo {pages}
  {46496} (\bibinfo {year} {2017})}\BibitemShut {NoStop}%
\bibitem [{\citenamefont {Puglisi}\ and\ \citenamefont
  {Marconi}(2017)}]{Puglisi2017}%
  \BibitemOpen
  \bibfield  {author} {\bibinfo {author} {\bibfnamefont {A.}~\bibnamefont
  {Puglisi}}\ and\ \bibinfo {author} {\bibfnamefont {U.~M.~B.}\ \bibnamefont
  {Marconi}},\ }\href@noop {} {\bibfield  {journal} {\bibinfo  {journal}
  {Entropy}\ }\textbf {\bibinfo {volume} {19}},\ \bibinfo {pages} {356}
  (\bibinfo {year} {2017})}\BibitemShut {NoStop}%
\bibitem [{\citenamefont {Sekimoto}(2010)}]{sekimoto_2010}%
  \BibitemOpen
  \bibfield  {author} {\bibinfo {author} {\bibfnamefont {K.}~\bibnamefont
  {Sekimoto}},\ }\href@noop {} {\emph {\bibinfo {title} {Stochastic
  Energetics}}}\ (\bibinfo  {publisher} {Springer},\ \bibinfo {address}
  {Berlin},\ \bibinfo {year} {2010})\BibitemShut {NoStop}%
\bibitem [{\citenamefont {Seifert}(2012)}]{Seifert_2012}%
  \BibitemOpen
  \bibfield  {author} {\bibinfo {author} {\bibfnamefont {U.}~\bibnamefont
  {Seifert}},\ }\href@noop {} {\bibfield  {journal} {\bibinfo  {journal}
  {Reports on Progress in Physics}\ }\textbf {\bibinfo {volume} {75}},\
  \bibinfo {pages} {126001} (\bibinfo {year} {2012})}\BibitemShut {NoStop}%
\bibitem [{\citenamefont {Andrieux}\ and\ \citenamefont
  {Gaspard}(2008)}]{andrieux2008quantum}%
  \BibitemOpen
  \bibfield  {author} {\bibinfo {author} {\bibfnamefont {D.}~\bibnamefont
  {Andrieux}}\ and\ \bibinfo {author} {\bibfnamefont {P.}~\bibnamefont
  {Gaspard}},\ }\href@noop {} {\bibfield  {journal} {\bibinfo  {journal}
  {Physical Review Letters}\ }\textbf {\bibinfo {volume} {100}},\ \bibinfo
  {pages} {230404} (\bibinfo {year} {2008})}\BibitemShut {NoStop}%
\bibitem [{\citenamefont {Pradhan}\ and\ \citenamefont
  {Seifert}(2010)}]{pradhan2010diamagnetism}%
  \BibitemOpen
  \bibfield  {author} {\bibinfo {author} {\bibfnamefont {P.}~\bibnamefont
  {Pradhan}}\ and\ \bibinfo {author} {\bibfnamefont {U.}~\bibnamefont
  {Seifert}},\ }\href@noop {} {\bibfield  {journal} {\bibinfo  {journal}
  {Europhysics Letters}\ }\textbf {\bibinfo {volume} {89}},\ \bibinfo {pages}
  {37001} (\bibinfo {year} {2010})}\BibitemShut {NoStop}%
\bibitem [{\citenamefont {Chun}\ and\ \citenamefont
  {Noh}(2018)}]{chun2018microscopic}%
  \BibitemOpen
  \bibfield  {author} {\bibinfo {author} {\bibfnamefont {H.-M.}\ \bibnamefont
  {Chun}}\ and\ \bibinfo {author} {\bibfnamefont {J.~D.}\ \bibnamefont {Noh}},\
  }\href@noop {} {\bibfield  {journal} {\bibinfo  {journal} {Journal of
  Statistical Mechanics: Theory and Experiment}\ }\textbf {\bibinfo {volume}
  {2018}},\ \bibinfo {pages} {023208} (\bibinfo {year} {2018})}\BibitemShut
  {NoStop}%
\end{thebibliography}

%
	
\end{document}